\documentstyle[preprint,aps]{revtex}

\begin{document}
\preprint{\begin{tabular}{c}
\hbox to\textwidth{December, 1994 \hfill SNUTP 94-94}\\[-10pt]
\hbox to\textwidth{hep-ph/9412397 }\\[36pt]
\end{tabular}}

\title{\bf Phenomenology of Soft Terms in the Presence
of Nonvanishing
Hidden Sector Potential Energy}
\author{Kiwoon Choi}
\address{\sl Department of Physics\\[-8pt]
Korea Advanced Institute of
Science and Technology\\[-8pt]
Taejon 305-701, Korea}

\author{Jihn E. Kim$^{(a,b)}$ and Gye T. Park$^{(b)}$}

\address{\sl
$^{(a)}$Department of Physics and
$^{(b)}$Center for Theoretical Physics\\[-8pt]
Seoul National University\\[-8pt]
Seoul 151-742, Korea}

\maketitle

\def\be{\begin{equation}}
\def\ee{\end{equation}}

\begin{abstract}
We argue that the hidden sector potential
energy is generically of order the intermediate scale
although the true cosmological
constant does vanish. This would significantly change the predicted values
of soft parameters for a variety of supergravity
models including those derived from string theory.
We stress that this point is particularly relevant for supergravity
models in which the soft masses
of observable fields vanish at tree level.
Implications of a nonzero
hidden sector potential energy
for low energy phenomenology are also studied.
\end{abstract}
\pacs{}

%\narrowtext

\noindent {\bf 1. Introduction}

With the accumulating precision data, and the prospect for
the operation of LEPII and upgraded Tevatron in the near future,
the phenomenological studies of models beyond the standard model
have become one of the major issues in particle physics.
The front runner among these models is the supergravity
models (SUGRA) \cite{rev}.  In SUGRA, the low energy effective
theory is described by supersymmetric terms plus soft supersymmetry
breaking terms of order $m_{3/2}$ which is of order
the electroweak scale.  For the electroweak scale phenomenology
in the minimal SUGRA,
there are five important parameters, the common scalar mass
$(m_0)$, the gaugino mass $(m_{1/2})$, $\tan\beta=v_2/v_1$, $A$,
and the top quark mass $m_{t}$.  $\mu$ and $B$ are determined
by the minimization condition.  With the anticipated confirmation
of the top quark mass at $m_t=174\pm 17$ GeV, there will be
four parameters in SUGRA phenomenology.  In addition, if supersymmetry
is broken by the $F$-component of the dilaton field in string theory, there
would exist two more relations defined {\it at the unification
scale}\footnote{Here the unification scale does not necessarily
mean the scale where the gauge coupling constants are unified.
It can be either  the Planck scale or the string
scale.}, $m_{1/2}=\sqrt{3}m_0$
and $A/\lambda=-\sqrt{3}m_0$, where
$\lambda$ denotes generic  Yukawa coupling constant
\cite{kl,barbieri}, while one has a different
boundary condition for a different route of supersymmetry breaking,
for example in no scale models
one would have  $m_0=0$ and $A=0$ at tree level
\cite{nano}.
Thus it is of utmost interest to figure out
what are the boundary values of soft parameters to get
an idea of the origin of the supersymmetry breaking.

As is well known, the soft scalar mass $m_0$ and the $B$ coefficient
at the unification scale depend on the value of
the hidden sector potential energy $V_h$\cite{ckn,brignole}.
When one computes
these soft parameters for a given SUGRA model, particularly
when one derives the above  mentioned relations between
soft parameters,
it has been  assumed that
$V_h$
does vanish\cite{rev}. This would be a consistent choice for the evaluation
of soft parameters {\it at tree level} since the tree level value
of the hidden sector potential energy corresponds to
the tree level cosmological constant.

However, in order to obtain a physically meaningful result
one needs to include  quantum effects.
In the usual procedure of taking quantum effects into account, one
considers renormalization group equations
in which logarithmically divergent corrections can be
properly included.  The boundary values
of the renormalized soft parameters at the unification scale
have been  still chosen to the values obtained
by assuming  the hidden sector potential energy to be zero.
However, it has been recently
pointed out \cite{ckn} that if one includes quantum corrections then
the hidden sector potential energy $V_h$ can be sizable,
large enough to significantly change the predicted
values of soft parameters, while the true
cosmological constant $V_{\rm eff}$ still does vanish.
The essential point was that, although they are the same at tree level,
$V_{\rm eff}$ and $V_h$ receive quantum corrections $\delta V_{\rm eff}$
and $\delta V_h$ which
are {\it completely independent} from each other.
In Ref. \cite{ckn}, an argument for a sizable $V_h$ was given
for models with
a large number of observable matter multiplets, $N=O(8\pi^2)$, whose
soft scalar masses are   comparable to the gravitino mass $m_{3/2}$.
In such models,  $\delta V_{\rm eff}$
would be dominated by
the quadratically divergent one loop contribution from the matter multiplets,
and then $\delta V_{\rm eff}\simeq Nm_0^2\Lambda^2/16\pi^2$ where $m_0$ denotes
the common soft scalar mass and $\Lambda$ is
the cutoff scale of the model\footnote{In Ref. \cite{ckn}, the whole
results were given in terms of
the three momentum cutoff $\Lambda_3$. Here we will use the more familiar
four momentum cutoff $\Lambda_4$. Note that $\Lambda_4^2=2\Lambda_3^2$.}.
 The hidden sector
potential energy $V_h$ also receives one loop correction, but
$\delta V_h$ is expected to be of
order $m_{3/2}^2\Lambda^2/16\pi^2$. Since we anticipate that our SUGRA
model is replaced by a well-behaved theory (for instance string theory)
around $\kappa^{-1}\equiv M_{Pl}/\sqrt{8\pi}$, the cutoff $\Lambda$
is expected to be of order $\kappa^{-1}$.
Then for $N=O(8\pi^2)$
and $m_0$ comparable to $m_{3/2}$,
the vanishing true cosmological constant, viz $V_0+\delta V_{\rm eff}=0$
where $V_0$ is the constant vacuum energy density at tree level,
implies that the magnitude  of $\kappa^2V_h$
can be comparable to   $m_0^2$. This means that
it is not a good approximation to ignore the $V_h$-dependent
part when one computes soft parameters for a given SUGRA model.
Note that here a nonzero $\kappa^2 V_h$ is essentially due
to quantum corrections, but it can be sizable because of the large
value of $N$. In this regard, one may recall that the conventional
logarithmic quantum corrections are important because
of the large value of $\ln(\Lambda^2/m_0^2)$.

The point that  the
$V_h$-dependent part of $m_0^2$ and $B$ can be sizable
is particularly relevant for
models in which the soft scalar masses
and the gaugino masses at the unification scale  {\it vanish at tree level}.
In fact, such a case occurs
quite often in string-inspired supergravity models\cite{brignole}.
In this case,  nonzero soft masses arise usually due
to string loop threshold effects. The resulting
soft masses are then suppressed by $g^2/16\pi^2$
compared to the gravitino mass or to the typical mass of
hidden sector fields, e.g. moduli fields,
triggerring  spontaneous supersymmetry breaking.
Then both the nonzero soft masses and
the nonzero $(V_h-V_{\rm eff})$ are the consequence of quantum
corrections at the same order, which means that
the $V_h$-dependent part of $m_0^2$ and $B$
is essentially the same order as the other parts.
Clearly then $V_h$-dependent part must be taken into account
for a consistent calculation of soft parameters.
Furthermore, as we will see later, the same argument holds true
even for models in which one loop quadratically divergent
vacuum energy density does vanish, whose motivation
has been recently discussed in Ref. \cite{zwirner,lopez}.

With the observation made above, one can introduce an additional
parameter $\epsilon$ which shows the importance of the $V_h$-dependent
part of soft parameters:
\be
\kappa^2 V_h\equiv -\epsilon m_0^2.
\ee
It becomes then necessary to know the value of $\epsilon$ in
order to get an idea of the
boundary values of the renormalized soft parameters.
In this paper, we wish to elaborate the arguments of Ref. \cite{ckn}
and study how sensitive are the low energy physics
at the electroweak scale to the presumably unknown parameter
$\epsilon$ defined in Eq. (1), mainly
focusing on string inspired supergravity models.

\noindent {\bf 2. Hidden Sector Potential Energy
and Expressions for Soft Parameters}

In this section, we elaborate  the arguments of Ref. \cite{ckn}
in the context of a  SUGRA model
with the following K\"{a}hler
potential and superpotential\footnote{In
the minimal supersymmetric standard model,
the only gauge and $R$-parity invariant term of the type
$\phi^i\phi^j$ is $\mu H_1H_2$.}
expanded in powers of the observable fields $\phi^i$ and
$\bar{\phi}^{\bar{i}}$:
\begin{eqnarray}
K&=&\tilde{K}(h,\bar{h})+ Z_{i\bar{j}}(h,\bar{h})\phi^i\bar{\phi}^{\bar{j}}+
\frac{1}{2}(Y_{ij}(h,\bar{h})\phi^i\phi^j+{\rm h.c.})+...,
\nonumber \\
W&=&\tilde{W}(h)+\frac{1}{2}\tilde{\mu}_{ij}(h)\phi^i\phi^j+
\frac{1}{3}\tilde{\lambda}_{ijk}(h)\phi^i\phi^j\phi^k+....
\end{eqnarray}
Here the hidden sector
fields triggering SUSY breaking are collectively denoted by
$h^{\alpha}$, and the observable sector fields $\phi^i$ include
the quarks, leptons, and the two Higgs doublets $H_1$ and $H_2$.

To obtain the effective action $S_{\phi}$ of the observable fields $\phi^i$,
we integrate out the hidden sector fields
\be
\exp (i S_{\phi}) = \int [{\cal D} h] \exp (i S),
\ee
where $S$ is the full supergravity action, and
$[{\cal D} h]$ includes the integration of the gravity multiplet
$(g_{\mu\nu}, \psi_{\mu})$ over
a background spacetime metric $\bar{g}_{\mu\nu}$ with macroscopic
wavelength. Since the high momentum modes of $\phi^i$ are
not integrated out yet, $S_{\phi}$ is defined at the cutoff
scale $\Lambda$ in the sense of Wilson. Thus to study
the low energy physics of $\phi^i$, one still needs to
scale the renormalization point down to the weak scale.
In the flat limit, $S_{\phi}$ would be characterized by
the effective superpotential of global SUSY \cite{rev}
\be
W_{\rm eff}\ =\ \frac{1}{3}\lambda_{ijk}\phi^i\phi^j\phi^k+\frac{1}{2}
\mu_{ij}\phi^i\phi^j,
\ee
and also  the soft breaking part  of the form
\be
{\cal L}_{\rm soft}\ =\ m^2_{i\bar{j}}\phi^i\bar{\phi}^{\bar{j}}+
(\frac{1}{3}A_{ijk}\phi^i\phi^j\phi^k+\frac{1}{2}B_{ij}\phi^i\phi^j+
\rm h.c.).
\ee

Supersymmetry breaking may have the seed in the hidden sector
gaugino condensation \cite{gcondensation}.  In this case, we
interpret our SUGRA model as the one obtained after integrating
out the hidden gauge multiplets (and also the
hidden gauge non-singlet matter multiplets if they exist).
Note that the hidden gauge multiplets can be integrated out
without breaking supersymmetry, and as a result  the effects of integration
can be summarized by the superpotential $\tilde{W}$ of the gauge singlet
hidden fields $h^{\alpha}$  whose nonzero $F$-components
are responsible for supersymmetry
breaking\footnote{Recently, there has been found a
discrepancy \cite{cm} between effective field theory
calculation \cite{bg} and the direct calculation \cite{cm} of
soft parameters in the gaugino condensation model
of supersymmetry breaking.  However, as pointed out in Ref. \cite{cm},
we believe that the two calculations should agree if one
includes the neglected terms of order $1/M_{Pl}^2$ in the
effective field theory framework.  In this case, supersymmetry
breaking by gaugino condensation can be also described by the
above superpotential given in Eq. (2).}.

Among the coefficients in $W_{\rm eff}$ and ${\cal L}_{\rm soft}$,
those which depend
on the hidden sector scalar potential $V_h$ are
relevant for us.
On dimensional ground,
we expect a correction of order
$\kappa^2 V_h$ for $m^2_{i\bar{j}}$ and $B_{ij}$
while
a correction of order $\kappa^3
V_h\ll m_{3/2}$  for $A_{ijk}$ and the gaugino masses.
(Here $\kappa=\sqrt{8\pi}/M_{Pl}$.)
Thus, for phenomenology, the
important correction can arise only in the soft
scalar masses and the $B$ terms.

To calculate soft terms in the scalar potential,
we first expand the supergravity
action in powers of the observable fields $\phi^i$.
Integrating out the hidden fields (see Eq. (3)), the
following soft scalar masses and the $B$ terms (for un-normalized
fields) are obtained\cite{kl}:
%\begin{equation}
%V\ =\ e^G\left[{\partial G\over \partial z^I}(G^{-1})^{I\bar{J}}{\partial G
%\over \partial \bar{z}^{\bar{J}} }-3\right]
%\end{equation}
%where
%\begin{equation}
%G\ =\ K+\log |W|^2,
%\end{equation}
%we obtain
\be
m^2_{i\bar{j}}\phi^i\bar{\phi}^{\bar{j}}+
\frac{1}{2}(B_{ij}\phi^i\phi^j+{\rm h.c.}),
%&&+[(1+|\xi|^2)m_{3/2}^2+\kappa_0^2V_h](H_1^*H_1+H_2^*H_2) \\
\ee
where\footnote{In fact, for a consistent study of quantum effects,
one needs to include in the expressions below
the part depending upon the hidden sector fermions. Here we will
ignore this point since our main point is {\it unchanged}.}
\begin{eqnarray}
m^2_{i\bar{j}} &&= \langle (\kappa^2V_h+m^2_{3/2})Z_{i\bar{j}}
-{F}^{\alpha}\bar{F}^{\bar{\beta}}R_{\alpha\bar{\beta} i \bar{j}})\rangle
\nonumber \\
&&=\langle \frac{2}{3}\kappa^2 V_h Z_{i\bar{j}}+
F^{\alpha}\bar{F}^{\bar{\beta}}(\frac{1}{3}K_{\alpha\bar{\beta}}Z_{i\bar{j}}
-R_{\alpha\bar{\beta}i\bar{j}})\rangle,
\nonumber \\
B_{ij} &&= \langle (\kappa^2 V_h+2m^2_{3/2}) Y_{ij}
+m_{3/2}(F^{\alpha}D_{\alpha}Y_{ij}-\bar{F}^{\bar{\alpha}}D_{\bar{\alpha}}
Y_{ij})-F^{\alpha}\bar{F}^{\bar{\beta}}D_{\alpha}D_{\bar{\beta}}Y_{ij}
\rangle
\nonumber \\
&&=\langle \frac{1}{3}\kappa^2 V_h Y_{ij}+m_{3/2}(F^{\alpha}
D_{\alpha}Y_{ij}-\bar{F}^{\bar{\alpha}}D_{\bar{\alpha}}Y_{ij})
+F^{\alpha}\bar{F}^{\bar{\beta}}(\frac{2}{3}K_{\alpha\bar{\beta}}-D_{\alpha}
D_{\bar{\beta}})Y_{ij}\rangle.
\end{eqnarray}
Here
we set $\tilde{\mu}_{ij}$ to zero since the
associated $B$-coefficients are not significantly affected
by the hidden sector potential energy, and
\begin{eqnarray}
&&\bar{F}^{\bar{\alpha}}
=e^{-i\theta}e^{\tilde{K}/2}\tilde{K}^{\beta\bar{\alpha}}
(\partial_{\beta}\tilde{W}+\tilde{W}\partial_{\beta}\tilde{K}),
\quad \theta ={\rm arg}(\tilde{W}),
\nonumber \\
&&D_{\alpha}Y_{ij}=\partial_{\alpha}Y_{ij}-\Gamma^l_{\alpha (i}Y_{j)l},
\quad  D_{\bar{\alpha}}Y_{ij}=\partial_{\bar{\alpha}}Y_{ij},
\nonumber \\
&&R_{\alpha\bar{\beta}i\bar{j}}=\partial_{\alpha}\partial_{\bar{\beta}}
Z_{i\bar{j}}-\Gamma^l_{\alpha i}\Gamma^{\bar{m}}_{\bar{\beta}\bar{j}}
Z_{l\bar{m}}, \quad \Gamma^l_{\alpha i}=Z^{l\bar{m}}\partial_{\alpha}
Z_{i\bar{m}},
\end{eqnarray}
and the gravitino mass $m_{3/2}$ and the hidden sector
scalar potential
$V_h$ are given by
\begin{eqnarray}
&&m_{3/2}=e^{\tilde{K}/2}|\tilde{W}|,
\nonumber \\
&&V_h=F^{\alpha}\bar{F}^{\bar{\beta}}\tilde{K}_{\alpha\bar{\beta}}-
3m^2_{3/2}.
\end{eqnarray}
%The other soft parameters and couplings are given by
%\begin{eqnarray}
%&&\lambda_{ijk}=\langle \tilde{\lambda}_{ijk}\exp (\kappa_0^2
%h_{\alpha}h^*_{\alpha}/2)\rangle,\nonumber \\
%&&\mu=
%\langle (\tilde{\mu}+\xi \kappa_0^2 W_h)
%\exp (\kappa_0^2 h_{\alpha} h^*_{\alpha}/2)\rangle , \\
%&&A_{ijk}=\langle \tilde{\lambda}_{ijk} \kappa_0^2 (h_{\alpha}
%D_{\alpha} W_h)^* \exp (\kappa_0^2 h_{\alpha} h^*_{\alpha})\rangle,
%\nonumber
%\end{eqnarray}
%where  $D_{\alpha}W_h=(\partial_{h_{\alpha}}+\kappa_0^2 h^*_{\alpha})W_h$.
The bracket means the average over the hidden
sector fields.  For example,
\be
\langle V_h\rangle =\int [{\cal D} h] V_h(h^{\alpha}) \exp (iS_h)/
\int [{\cal D} h] \exp (i S_h),
\ee
where $S_h$ is the supergravity action of the hidden sector fields alone.

Eq. (7) shows that $m^2_{i\bar{j}}$ and $B_{ij}$
associated with the K\"{a}hler
potential term $Y_{ij}\phi^i\phi^j$\cite{masiero}
depend on $\langle V_h Z_{i\bar{j}}\rangle$ and $\langle V_h Y_{ij}\rangle$
respectively\cite{ckn,brignole}.
In the following, we assume the factorization approximation
is valid  and thus
\be
\langle V_h Z_{i\bar{j}}\rangle \simeq
\langle V_h\rangle \langle Z_{i \bar{j}}
\rangle, \quad
\langle V_h Y_{ij} \rangle \simeq \langle V_h \rangle \langle Y_{ij}
\rangle.
\ee
For local SUSY broken by a nonzero value
of the auxiliary component $F^{\alpha}$, {\it unless} one implements
a fine tuning of  some parameters in the hidden sector
superpotential $\tilde{W}$, the typical size of $\langle V_h\rangle$
would be of $O(|F^{\alpha}|^2)=O(\kappa^{-2} m_{3/2}^2)$.
In most of the previous studies, motivated by the vanishing
cosmological constant, the expectation value of the hidden sector
scalar potential $\langle V_h\rangle$ was simply assumed to be zero.
However, as we argue below, $\kappa^2\langle V_h\rangle$
is generically  of order the soft scalar mass squared
of observable fields, and thus ignoring the
$\langle V_h\rangle$-dependent part of soft parameters is {\it not}
a sensible approximation.

In this regard, a simple but essential point is that
$\langle V_h\rangle$ is not the true cosmological constant.
The fully renormalized cosmological constant
$V_{\rm eff}$ at low energy is obtained by integrating
out all the fields in the theory,\footnote{The long wave length
metric $\bar g_{\mu\nu}$ is treated as background and is
not integrated out to see the gravitational effect at low energy.}
\be
\exp (i\int d^4 x \sqrt{\bar g} V_{\rm eff})=\int [{\cal D}\phi {\cal D}h]
\exp (i\int d^4 x \sqrt{g}{\cal L})
\ee
where $[{\cal D}\phi]$ represents the integration over all
the observable gauge and matter multiplets.
In the classical approximation, both $\langle V_h \rangle$ of Eq. (10)
and $V_{\rm eff}$ of  Eq. (12) are simply the classical potential
in $S_h$.
Therefore, we have
\be
(V_{\rm eff})_{\rm tree}=\langle V_h\rangle_{\rm tree}.
\ee
However, beyond the tree level, obviously $\langle V_h\rangle$
is {\it not} the same as the fully renormalized
cosmological constant.
For instance, $\langle V_h\rangle$ of Eq. (10) is completely
independent of the observable sector dynamics, while $V_{\rm eff}$
of Eq. (12) depends on.

To proceed, let us consider a toy model which is calculable
and at the same time shows the point essential to us.
The model contains  a single hidden chiral field $h$ and $N$
observable chiral fields $\phi^i$ with the following lagrangian:
\begin{eqnarray}
{\cal L}_{\rm toy} &=& \partial_{\mu}h\partial^{\mu}
h^*+\partial_{\mu}\phi^i
\partial^{\mu}\phi^{i*}+\frac{1}{2}\bar{\psi}(i\gamma^{\mu}
\partial_{\mu}-m_F)\psi+\frac{1}{2}\bar{\chi}^ii\gamma^{\mu}
\partial_{\mu}\chi^i
\nonumber \\
&&-[V_0+m_B^2hh^*+\{\kappa^2(V_0+m_B^2hh^*)+\bar{m}_0^2\}\phi^i\phi^{i*}]
,
\end{eqnarray}
where $\psi$ and $\chi^i$ denote the superpartner Majorana fermions of $h$
and $\phi^i$, respectively. Here the theory
is regulated by an explicit cut-off scale $\Lambda$ which
is of order $\kappa^{-1}=M_{Pl}/\sqrt{8\pi}$.
The mass parameters $m_B$, $m_F$ and $\bar{m}_0$ are
of order the electroweak scale,
and the constant vacuum energy density $V_0$ can be
of $O(\kappa^{-2}\bar{m}_0^2 )$.  In this toy model,
supersymmetry appears to be explicitly broken for  nonzero
values of the parameters $\kappa^2V_0$, $(m_B^2-m_F^2)$, $\bar{m}_0^2$,
and $m_B^2$ which are introduced
to mimic the potentially complicated hidden sector dynamics
which would trigger  spontaneous supersymmetry breaking.
Thus by definition these supersymmetry breaking parameters
can {\it not} be significantly larger than $m^2_{3/2}$.

The hidden sector potential energy of our toy model is
\be
V_h=V_0+m_B^2 hh^*,
\ee
and the soft mass of the observable $\phi^i$  is
\be
m_0^2=\bar{m}_0^2+\kappa^2{\langle V_h\rangle}.
\ee
Obviously at tree level,
\be
(V_{\rm eff})_{\rm tree}=\langle V_h\rangle_{\rm tree}=V_0.
\ee
Using Eqs. (10) and (12),
one can easily compute the cosmological constant $V_{\rm eff}$ and
the expectation value of the hidden sector potential
energy $\langle V_h\rangle$
at the full quantum level. The results are
\begin{eqnarray}
&& \langle V_h\rangle =V_0+\frac{1}{16\pi^2}m_B^2\Lambda^2,
\nonumber \\
&& V_{\rm eff}=V_0+V_1+V_2+O(m_B^4),
\end{eqnarray}
where $V_1$ and $V_2$ denote the one loop and the two loop
corrections to the vacuum energy density:
\begin{eqnarray}
V_1 &= & \frac{1}{16\pi^2}\Lambda^2[(m_B^2-m_F^2)+
N(\bar{m}_0^2+\kappa^2V_0)],
\nonumber \\
V_2 &=&  \frac{N}{16\pi^2}\left(\frac{\kappa^2 \Lambda^2}{16\pi^2}\right)
m_B^2\Lambda^2 .
\end{eqnarray}
Note that the above result of $\langle V_h\rangle$ is exact
and also that of $V_{\rm eff}$ is exact up to small corrections
of $O(m_B^4)$.

The results of Eqs. (18) and (19)  show apparently that the quantum
correction to $\langle V_h\rangle$ and the correction
to $V_{\rm eff}$ are completely independent from each other.
Imposing the condition of vanishing cosmological constant, viz
$V_{\rm eff}=0$, we find
\begin{eqnarray}
\kappa^2\langle V_h\rangle &=& - \frac{\kappa^2\Lambda^2}{16\pi^2}[N(
\bar{m}_0^2+\kappa^2 V_0)+\frac{N\kappa^2\Lambda^2}{16\pi^2})m_B^2-
m_F^2], \nonumber \\
&=& -\frac{\kappa^2\Lambda^2}{16\pi^2}[Nm_0^2-m_F^2]
\nonumber \\
&\equiv & -\epsilon m_0^2.
\end{eqnarray}

Using the above results of  our toy model,
we can discuss the size of $\epsilon$
for a variety of cases. Before going further, let us note
that our toy model does not contain gauge multiplets.
In realistic models, the quantum corrections
to $V_{\rm eff}$ would receive an additional
contribution from the observable sector
gaugino masses, while $\langle V_h\rangle$ is not affected by the
observable sector dynamics. However, in most interesting SUGRA models
the gaugino mass $m_{1/2}$ is either comparable to
or significantly smaller than $m_0$.
For $m_{1/2}$ comparable to $m_0$, we will assume
that the number of chiral matter multiplets is significantly
greater than that of gauge multiplets.  Then the following
discussion in the context of our toy model
will be applied also for realistic models.

Let us first consider the case that all of
$\bar{m}_0$, $m_B$, and $m_F$ are comparable to $m_{3/2}$.
(Note that $\bar{m}_0$ and $m_0$ denote the tree level
mass and the quantum  corrected  mass
of observable scalar fields at the unification scale, while
$m_B$ and $m_F$ denote the typical mass of the hidden sector
scalar and the hidden sector fermion, respectively.)
The dilaton dominated scenario in string theory
and also generic SUGRA models with the flat K\"{a}hler potential
would correspond to this case.  In this case, Eq. (20) implies
$\epsilon=O(N\kappa^2\Lambda^2/16 \pi^2)$.  Then  $\epsilon$
can be significant  if the cutoff\footnote{
Here the cut off scale $\Lambda$ would not be exactly the same as the
frequently quoted mass scale $M_{\rm GUT}=e^{(1-\gamma)/2} 3^{-3/4}g
\kappa^{-1}/\sqrt{2\pi}$ in string theory\cite{kaplunovsky}
though both $\Lambda$ and $M_{\rm GUT}$
are comparable to the string scale
$M_{\rm st}=g\kappa^{-1}$. Note that a precise
definition of $M_{\rm GUT}$ is a matter of convention and
the above choice was made for a particular way to incorporate
the threshold correction to the renormalized  gauge
coupling constants\cite{kaplunovsky}. }
is comparable to $\kappa^{-1}$
and $N=O(8\pi^2)$. Furthermore, in this case,
it is likely that $\epsilon$ is positive\cite{ckn}.

Even more interesting case is  the one
in which  $m_0^2$ is of order $g^2m_{3/2}^2/16\pi^2$,
while $m_B$ and $m_F$ are still comparable to $m_{3/2}$.
This would be the case for many of the moduli dominated
supersymmetry breaking
scenarios in string theory which give  vanishing soft scalar and
gaugino masses  of observable fields {\it at tree level}.
In such cases, soft masses are induced by string loop
threshold corrections\cite{brignole}.
The resulting $m_0/m_{3/2}$ and $m_{1/2}/m_{3/2}$
are of order $\sqrt{g^2/16\pi^2}$ and $g^2/16\pi^2$,
respectively, implying that $m_{1/2}$ is significantly
smaller than $m_0$.  Then Eq. (20) gives
$\epsilon=O(\frac{\kappa^2\Lambda^2}{16\pi^2}
\frac{16\pi^2}{g^2})$. This shows that
$\epsilon$ does {\it not} represent a correction
to the leading result, but rather it
corresponds to the ratio between two independent one loop
effects and thus of order unity in general.

Recently an interesting class of string inspired no scale SUGRA models
have been considered  in Ref. \cite{zwirner,lopez}.
The peculiar property of those models
was that both the tree level vacuum energy density and the one
loop quadratically divergent vacuum energy density
vanish, viz $V_0=V_1=0$.
Typically such models also have the vanishing soft masses
of observable
fields at tree level, viz $\bar{m}_0=0$,
 while some of hidden sector fields have masses
of order $m_{3/2}$.
Then with $\bar{m}_0=V_0=V_1=0$, we find
\be
\langle V_h\rangle =
\frac{\kappa^2\Lambda^2}{16\pi^2}(1-\frac{N\kappa^2\Lambda^2}{16\pi^2})m_B^2
\equiv -\epsilon m_0^2.
\ee
Again with $m_0^2=O(g^2m_{3/2}^2/16\pi^2)$ induced by string loop
threshold and $m_B=O(m_{3/2})$,
we find $\epsilon$ is essentially of order unity in this case also\footnote{
In the models with $V_0=V_1=0$ which have been considered
in Ref. \cite{zwirner}, some of hidden scalars are extremely
light, having masses
of order $\kappa m_{3/2}^2$. However our argument here is valid as long as
there exists any  hidden scalar with its mass $m_B=O(m_{3/2})$,
which is always the case for the models of Ref. \cite{zwirner}.}.

So far, we have argued that
the boundary values of
scalar mass $m_0^2$ and the $B$ parameter in generic SUGRA
models can be significantly
changed compared to those in the naive approach assuming the
hidden sector potential $V_h$ to zero.
Thus when one computes
soft parameters for a given SUGRA model, one needs
to introduce an additional parameter $\epsilon$
which is defined by $\langle V_h\rangle\equiv -\epsilon m_0^2$
to parameterize the importance of the contribution
from  nonzero hidden sector potential energy.
In most of interesting cases, $\epsilon$ could be (or was
essentially) of order unity.
We wish to stress again that for the case that the soft masses
of observable sector vanish at tree level, which
is the case that occurs
quite often in string inspired SUGRA, $\epsilon$ does {\it
not} represent a correction to the leading results,
but rather corresponds to the  ratio between two independent
one loop effects.
Then theories predicting soft parameters would suffer from
an uncertainty associated with the potentially sizable value
of $\epsilon$.
In the next section, we investigate how sensitive the low
energy phenomenology is to the parameter $\epsilon$.

\noindent {\bf 3. The $\epsilon$ Dependence of Masses of
Superpartners at Low Energy}

In this section, we will
evolve the $\epsilon$-dependent boundary values to the
electroweak scale to see how sensitive
the low energy physical parameters are to the  value
of $\epsilon$. To be explicit, we consider
two specific scenarios for supersymmetry breaking which
lead to very distinctive predictions of soft parameters:
the dilaton-dominated model $\cite{kl,barbieri}$  and the moduli-dominated
orbifold model with small string loop threshold corrections
and all modular weights of matters fields being $-1$ $\cite{brignole}$.

For a numerical study,
we assume the minimal particle content
in the observable sector, viz.
the particles of the minimal supersymmetric standard model (MSSM).
If there exist more particles, with masses between
the unification scale  and the electroweak scale,
transforming nontrivially under $SU(3)\times
SU(2)\times U(1)$, our results would be changed accordingly.
For our purpose, it is sufficient to consider the
following renormalization group equations $\cite{rge}$ keeping only
the leading terms in the mass hierarchy in the three generation MSSM.
\begin{eqnarray}
{{dM_a}\over {dt}}&=&{2\over {16\pi ^2}}b_ag_a^2M_a\;, \nonumber \\
{{dA_t}\over {dt}}&=&{2\over {16\pi ^2}}\Big (\sum c_ag_a^2M_a
+6\lambda _t^2A_t+\lambda _b^2A_b\Big )\;, \nonumber \\
{{dA_b}\over {dt}}&=&{2\over {16\pi ^2}}\Big (\sum c_a^{\prime }g_a^2M_a
+6\lambda _b^2A_b+\lambda _t^2A_t+\lambda _{\tau}^2A_{\tau}\Big )\;,
\nonumber \\
{{dA_{\tau }}\over {dt}}&=&{2\over {16\pi ^2}}\Big (\sum c_a^{\prime \prime }
g_a^2M_a+3\lambda _b^2A_b+4\lambda _{\tau}^2A_{\tau}\Big )\;, \nonumber \\
{{dB}\over {dt}}&=&{2\over {16\pi ^2}}\Big ({3\over 5}g_1^2M_1+3g_2^2M_2
+3\lambda _b^2A_b+3\lambda _t^2A_t+\lambda _{\tau}^2A_{\tau}\Big )\;,
\nonumber \\
{{d\mu }\over {dt}}&=&{\mu \over {16\pi ^2}}\Big (-{3\over 5}g_1^2-3g_2^2
+3\lambda _t^2+3\lambda _b^2+\lambda _{\tau}^2\Big )\;, \nonumber \\
{{dM_{H_1}^2}\over {dt}}&=&{2 \over {16\pi ^2}}
\Big (-{3\over 5}g_1^2M_1^2-3g_2^2M_2^2
+3\lambda _b^2X_b+\lambda _{\tau}^2X_{\tau }\Big )\;, \nonumber \\
{{dM_{H_2}^2}\over {dt}}&=&{2 \over {16\pi ^2}}
\Big (-{3\over 5}g_1^2M_1^2-3g_2^2M_2^2
+3\lambda _t^2X_t\Big )\;, \nonumber \\
{{dM_{Q_L}^2}\over {dt}}&=&{2 \over {16\pi ^2}}
\Big (-{1\over 15}g_1^2M_1^2-3g_2^2M_2^2-{16\over 3}g_3^2M_3^2
+\lambda _t^2X_t+\lambda _b^2X_b\Big )\;, \nonumber \\
{{dM_{t_R}^2}\over {dt}}&=&{2 \over {16\pi ^2}}
\Big (-{16\over 15}g_1^2M_1^2-{16\over 3}g_3^2M_3^2
+2\lambda _t^2X_t\Big )\;,  \nonumber \\
{{dM_{b_R}^2}\over {dt}}&=&{2 \over {16\pi ^2}}
\Big (-{4\over 15}g_1^2M_1^2-{16\over 3}g_3^2M_3^2
+2\lambda _b^2X_b\Big )\;, \nonumber \\
{{dM_{L_L}^2}\over {dt}}&=&{2 \over {16\pi ^2}}
\Big (-{3\over 5}g_1^2M_1^2-3g_2^2M_2^2
+\lambda _{\tau}^2X_{\tau }\Big )\;, \nonumber \\
{{dM_{\tau _R}^2}\over {dt}}&=&{2 \over {16\pi ^2}}
\Big (-{12\over 5}g_1^2M_1^2
+2\lambda _{\tau}^2X_{\tau }\Big )\;,
\end{eqnarray}
%\end{mathletters}
%
and for the two light generations,
%
%\begin{mathletters}
\begin{eqnarray}
{{dA_u}\over {dt}}&=&{2\over {16\pi ^2}}\Big (\sum c_ag_a^2M_a
+\lambda _t^2A_t\Big )\;, \nonumber \\
{{dA_d}\over {dt}}&=&{2\over {16\pi ^2}}\Big (\sum c_a^{\prime }g_a^2M_a
+\lambda _b^2A_b+{1\over 3}\lambda _{\tau}^2A_{\tau}\Big )\;, \nonumber \\
{{dA_e}\over {dt}}&=&{2\over {16\pi ^2}}\Big (\sum c_a^{\prime \prime }
g_a^2M_a+\lambda _b^2A_b+{1\over 3}\lambda _{\tau}^2A_{\tau}\Big )\;,
\nonumber \\
{{dM_{q_L}^2}\over {dt}}&=&{2 \over {16\pi ^2}}
\Big (-{1\over 15}g_1^2M_1^2-3g_2^2M_2^2-{16\over 3}g_3^2M_3^2
\Big )\;, \nonumber \\
{{dM_{u_R}^2}\over {dt}}&=&{2 \over {16\pi ^2}}
\Big (-{16\over 15}g_1^2M_1^2-{16\over 3}g_3^2M_3^2
\Big )\;, \nonumber \\
{{dM_{d_R}^2}\over {dt}}&=&{2 \over {16\pi ^2}}
\Big (-{4\over 15}g_1^2M_1^2-{16\over 3}g_3^2M_3^2
\Big )\;, \nonumber \\
{{dM_{l_L}^2}\over {dt}}&=&{2 \over {16\pi ^2}}
\Big (-{3\over 5}g_1^2M_1^2-3g_2^2M_2^2
\Big )\;, \nonumber \\
{{dM_{e_R}^2}\over {dt}}&=&{2 \over {16\pi ^2}}
\Big (-{12\over 5}g_1^2M_1^2
\Big )\;,
\end{eqnarray}
%\end{mathletters}
%
where $M_a\ (a=1,2,3)$ are the gaugino masses of $SU(3)\times
SU(2)\times U(1)$, $M_\phi$ are the scalar masses, and
%
%\begin{mathletters}
\begin{eqnarray}
b_a&=&({33\over 5},1,-3) \;, \nonumber \\
c_a&=&({13\over 15},3,{16\over 3}) \;, \nonumber \\
c_a^{\prime}&=&({7\over 15},3,{16\over 3}) \;, \nonumber \\
c_a^{\prime \prime}&=&({9\over 5},3,0) \;, \nonumber \\
X_t      &=&  M_{Q_L}^2+M_{t _R}^2+M_{H_2}^2+A_t^2\;, \nonumber \\
X_b      &=&  M_{Q_L}^2+M_{b _R}^2+M_{H_1}^2+A_b^2\;, \nonumber \\
X_{\tau }&=&  M_{L_L}^2+M_{\tau _R}^2+M_{H_1}^2+A_{\tau }^2\;.
\end{eqnarray}
%\end{mathletters}
%
Here, $t=\ln (P/M_U)$ for the renormalization
point $P$ and the unification scale $M_U$,
and the factors
$c_a$, $c_a^{\prime }$, and $c_a^{\prime\prime  }$ are
given by a sum over the fields in the relevant Yukawa coupling,
e.g. $c_a=\sum _f c_a(f)=c_a({H_2})+c_a({Q})+c_a({U^c})$.  The Yukawa
couplings are defined by
\be
\lambda_t={\sqrt{2}m_t\over v\sin \beta},\ \
\lambda_b={\sqrt{2}m_b\over v\cos\beta}, \ \
\lambda_\tau ={\sqrt{2}m_\tau\over v\cos\beta}.
\ee

Before performing a numerical analysis, let us
summarize the relevant formulae of soft parameters
which have been derived from string inspired supergravity models.
Following  Brignole, Ibanez and Munoz \cite{brignole},
we parametrize the soft parameters at the unification scale
in the following manner,
\begin{eqnarray}
m_0^2\ &=&\ \kappa^2V_h+m_{3/2}^2-\frac{1}{3}\kappa^2|F|^2(1-\delta_0)
\cos^2\theta \nonumber \\
&=&\ {1\over 3}\kappa^2\left[2V_h+|F|^2(\sin^2\theta+\delta_0\cos^2\theta)
\right], \nonumber \\
M_a\ &=&\ (1+\eta_a)\kappa F(\sin\theta+\delta_a\cos\theta), \nonumber \\
A\ &=&\ -\kappa F\sin\theta,
\end{eqnarray}
where $\kappa=\sqrt{8\pi}/M_{Pl}$,
the Goldstino  angle $\theta$ is defined by $\tan\theta=F_S/F_T$, the small
parameters $\delta_0$, $\delta_a$, and $\eta_a$ represent
string loop effects,
$|F|^2\ =\ F^{\alpha}\bar{F}^{\bar{\beta}}K_{\alpha\bar{\beta}}$
corresponds to the scale of local supersymmetry breaking,
and finally
the hidden sector potential energy  is given by
$V_h=|F|^2-3\kappa^{-2}m_{3/2}^2$.
Note that $F$, $V_h$, and $\theta$ are determined by
nonperturbative dynamics, i.e. highly depend on
the hidden sector superpotential $\tilde{W}(h)$ which is presumed
to be generated by nonperturbative effects,
while $\delta_0, \delta_a$, and $\eta_a$
are perturbative parameters which can be calculated in string
perturbation theory.
The parametrization given above is valid in fact for a wide class of
superstring models, e.g. large size limit of Calabi-Yau manifolds,
generic dilaton dominated cases, and orbifold models  with
small string threshold corrections and all modular weights of matter
fields being $-1$.

Then the parametrization taking the possibility of significantly
large $\kappa^2V_h=-\epsilon m_0^2$ into account
is given by
\begin{eqnarray}
m_0^2\ &=& \ \frac{1}{3}(1+{2\over 3}\epsilon)^{-1}\kappa^2 |F|^2
(\sin^2\theta+\delta_0\cos^2\theta), \nonumber \\
M_a\ &=& \ (1+\eta_a)\kappa F(\sin\theta+\delta_a\cos\theta), \nonumber\\
A\ &=&\ -\kappa F\sin\theta.
\end{eqnarray}

{\it
(i) Dilaton--dominated case with small string-loop threshold
effect}\,: \, Here we assume $\sin^2\theta\gg\delta_0,\delta_a, \eta_a$,
and thus
\begin{eqnarray}
m_0^2\ &\simeq & {1 \over 3}(1+{2\over 3}\epsilon)^{-1}
\kappa^2 |F|^2
\sin^2\theta ={1 \over 3 +2\epsilon}m_{1/2}^2, \nonumber
\\
M_a\ &\simeq&\ \kappa F\sin\theta= m_{1/2}, \nonumber \\
A\ &=&\ -\kappa F\sin\theta=-m_{1/2}.
\end{eqnarray}

{\it (ii) Moduli--dominated orbifold case
with the modular weights of all matters being $-1$}\,:
Here we assume $\delta_{GS}=10$ and $\sin\theta =0$, and then
\begin{eqnarray}
\delta_0&=&{g^2\over 48\pi^2}\delta_{GS}
=10^{-2}, \nonumber \\
\eta_0&=&0\ ,\ \ \eta_2 =0.06\ ,\ \ \eta_3=0.18\ , \nonumber \\
\delta_3&=& (\delta_{GS}-3)\times 4.6\times 10^{-4}=3.22\times 10^{-3},
\nonumber \\
\delta_2&=& (\delta_{GS}+1)\times 4.6\times 10^{-4}=5.06\times 10^{-3},
\nonumber\\
\delta_1&=& (\delta_{GS}+{33\over 5})\times 4.6\times 10^{-4}
=7.64\times 10^{-3},
\end{eqnarray}
which give
\begin{eqnarray}
m_0^2 &=&
3.33 \times 10^{-3} (1+{2\over 3}\epsilon)^{-1}\kappa^2 |F|^2, \nonumber \\
M_3 &=& 3.22\times 10^{-3}\kappa F, \nonumber \\
M_2 &=& 5.36\times 10^{-3} \kappa F, \nonumber \\
M_1 &=& 9.02\times 10^{-3}\kappa F, \nonumber\\
A &=& 0.
\end{eqnarray}

The soft supersymmetry breaking parameters relevant for
$SU(2)\times U(1)$ breaking are $m_0^2$, $m_t$, $A$, $B$,
and the gaugino masses.
Also the supersymmetric Higgs mixing
parameter $\mu$ contributes to the $SU(2)\times U(1)$ breaking.
With these parameters, we search for the mimimum of the
potential determining $\langle H_1^0\rangle=v_1$ and $\langle H_2^0
\rangle=v_2$.  We then trade
$\mu$ and $B$ for $\tan\beta=v_2/v_1$ and $v=\sqrt{v_1^2
+v_2^2}$.
In the presence of a nonzero $\kappa^2V_h=-\epsilon m_0^2$,
only $m_0^2$ and $B$ can be significantly affected.
In our numerical analysis for the boundary conditions of
Eqs. (28) and (30),
only the $\epsilon$-dependence of $m_0^2$
has been implemented, while $B$ is
determined by imposing the radiative electroweak symmetry
breaking.

Let us first summarize the results
of the dilaton-dominated case ({\it i}). As was explained in
the previous section, in this case $\epsilon$ would represent
a potentially large quantum correction associated with
the large value of $N$, the number of observable chiral
multiplets. In Fig. 1,
we present the masses of the third generation squarks and sleptons
at the electroweak scale for
a particular choice of parameters with
$m_{1/2}=150$ GeV, $m_t=170$ GeV, and $\tan\beta=2$.
There $\epsilon$ is allowed to vary from zero to one.
It shows that the effect of a nonzero hidden sector potential
energy for the low energy squark masses
is negligible in this case.
The reason for this can be traced back to the renormalization
group equations for squarks in Eqs. (22) and (23)
which show that, for the gluino mass bigger than the squark mass,
the low energy squark masses
are determined mainly by the $\epsilon$-independent
$g_3M_3$, {\it not} by the $\epsilon$-dependent boundary value
of $m_0^2$ at the unification scale.
Therefore, changing the boundary value of $m_0^2$ by order unity
does not change the low energy squark masses very much.
On the other hand, the change of the slepton masses is
somewhat noticeable.  For example, for $\epsilon=0.6$ the slepton
masses decrease by 8 \%. The renormalization group
equation for sleptons still contains the contribution from
$g_2M_2$, but its effect
is not as dramatic as that of the squark case since it involves
the weak gauge coupling constant $g_2$.
This can be more easily understood from the
anaytic expressions for squark and slepton masses in the
two light generations given by
\be
M^2_{q_L}\simeq m_0^2+8m_{1/2}^2\ ,\ \ \ M^2_{l_L}
\simeq m_0^2+0.63 m_{1/2}^2\ ,
\ee
where the effect of changing $m_0$ through $\epsilon$ in
$M^2_{q_L}(M^2_{l_L})$ becomes negligible (pronounced)
because of a large (small) coefficient of $m^2_{1/2}$.
In Fig. 2, we present the first generation
squark masses as a function of $\epsilon$.  In Fig. 3, we present
the Higgs boson masses as a function of $\epsilon$.  The Higgs boson
masses are insensitive to $\epsilon$.
Similarly in Fig. 4, we show chargino
($\chi^+_i$) and neutralino ($\chi^0_i$) masses.
In summary, for the dilaton dominated case ({\it i}),
except for the slepton masses, corrections to low energy soft parameters
due to $\epsilon$ are not significant at all.

Let us now consider Case ({\it ii}) of moduli-dominated scenario.
As was stressed, $\epsilon$ in this case corresponds to the ratio
between two independent one loop effects, and thus is essentially
of order unity. Thus compared to the case ({\it i}),
$\epsilon$ can vary over a wider range, but we restrict it up to
1 to compare with Figs. (1)--(4).  (However, keep in mind that
$\epsilon$ can take values greater than 1.)
Furthermore,  gaugino masses in Case ({\it ii}) are significantly
smaller than the soft scalar masses.  As a result,
low energy soft parameters would be more sensitive
to the $\epsilon$-dependent boundary value of $m_0^2$.
In Fig. 5, we present the $\epsilon$ dependence of the
third generation squark and slepton masses.  Here the input
masses of $m_{1/2}=100$ GeV, $m_t=150$ GeV, and $\tan\beta=2$
are used.  $m_{1/2}$ is the gluino mass at the unification
scale and the wino and bino masses at the unification scale
are given by Eq. (30).  As expected, the moduli dominated scenario
has the stronger $\epsilon$ dependence.  For example, the third
generation slepton masses decrease by $\sim 15$ \% when one varies
$\epsilon$ from 0 to 0.6.  In Fig. 6, we present the first
generation squark masses.  Here also, the $\epsilon$ dependence
of the masses are pronounced.  In Figs. 7 and 8, we present
the Higgs boson masses and chargino and neutralino masses,
respectively.

\noindent {\bf 4. Conclusion}

In the previous studies of soft parameters within
supergravity models, the hidden sector potential energy
has been usually assumed to vanish.  The only motivation
for this assumption is the vanishing cosmological constant.
In this paper we expanded the discussion of Ref. \cite{ckn}
that although they are the same at tree level
the vacuum energy density $V_{\rm eff}$
and the hidden sector potential energy $V_h$ receive
completely different quantum corrections.
As shown for the simple model in Sec. 2,
$V_h$ (the hidden sector vacuum energy) is
generally of order $\kappa^{-2}m_0^2\equiv m_0^2 M_{Pl}^2/8\pi$
for $V_{\rm eff}$ (the true cosmological constant) = 0,
where $m_0$ denotes the soft mass of observable sector
scalar fields at the unification scale.
This leads to a physically interesting consequence in particle physics
since $m_0^2$ and $B$ depend on $\kappa^2V_h$.
The value of $\kappa^2V_h$ of order $m_0^2$ gives rise to
a significant change in the predicted values of soft
parameters in a variety of supergravity models.

It should be stressed that the above point is particularly
relevant for supergravity models in which  the soft masses
of observable fields vanish at tree level, which is the case
occurring quite often in string inspired supergravity models
\cite{brignole}.  In such models, both nonzero $m_0^2$  and
nonzero $(V_h-V_{\rm eff})$ are the consequence
of quantum effects at the same order. Then $\epsilon=-\kappa^2 V_h/m_0^2$
corresponds to the ratio of two independent one loop effects,
and thus is essentially of order unity.
The same observation holds true for the recently discussed no scale
models \cite{zwirner,lopez} in which both the tree level vacuum
energy density and the one loop quadratically divergent vacuum
energy density vanish.

In order to see the low energy implication of the nonvanishing
$V_h$, we have performed the renormalization group
evolution starting from the $\epsilon$-dependent $m_0^2$
at the unification scale.  We have imposed explicitly
the radiative electroweak symmetry breaking.
In the dilaton dominated supersymmetry
breaking scenario, the slepton masses have a sizable
$\epsilon$ dependence, but the other masses are insensitive to
$\epsilon$.  In the moduli dominated scenario, the $\epsilon$
dependence of soft masses are much more pronounced.

\acknowledgments
This work is supported in part by the Korea Science and Engineering
Foundation through Center for Theoretical Physics, Seoul
National University (JEK,GTP), KOSEF--DFG Collaboration Program (JEK),
the Basic Science Research Institute Program, Ministry of
Education, 1994,  BSRI-94-2418 (JEK) and BSRI-94-2434 (KC).

\begin{figure}
\caption{The third generation squark [(a) $\tilde t$,
$\tilde b$] and slepton [(b) $\tilde \nu$, $\tilde \tau$] masses
as a function of $\epsilon$  in the dilaton dominated
supersymmetry breaking scenario.  Stop, sbottom and gluino are
represented by $\tilde t$, $\tilde b$, and $\tilde g$, respectively.
The masses are in units of GeV.
Among two mass eigenstates of $\tilde t$, $\tilde b$ and
$\tilde\tau$, the heavier ones
are suffixed with 2 and the lighter ones with 1.
We use the following set of parameters,
$m_{\tilde g}=325$ GeV, $m_t=170$ GeV, and $\tan\beta=2$.
These masses are the values at the electroweak scale. The
universal gaugino mass of 150 GeV is given at the unification scale.}
\end{figure}
\begin{figure}
\caption{The first generation squark ($\tilde u$, $\tilde d$)
masses as a function of $\epsilon$ in the dilaton scenario.
The input parameters are the same as in the Fig.~1. We
do not present the first generation slepton masses, since they are
almost the same as the third generation slepton masses.}
\end{figure}
\begin{figure}
\caption{The Higgs boson masses as a function of $\epsilon$
in the dilaton scenario. The lightest Higgs boson is $h$,
and the neutral pseudoscalar Higgs boson is $A$.
The input parameters are the same as in the Fig.~1.}
\end{figure}
\begin{figure}
\caption{Same as in the Fig.~1 except that
the chargino and neutralino masses are presented.}
\end{figure}
\begin{figure}
\caption{The third generation squark and slepton masses as
a function of $\epsilon$ in the moduli dominated supersymmetry
breaking scenario. The input parameters are
$m_{1/2}=100$ GeV, $m_t=150$ GeV, and $\tan\beta=2$.
$m_{1/2}$ is the gluino mass at the unification scale. The
wino and bino masses at the unification scale are given by Eq. (30).}
\end{figure}
\begin{figure}
\caption{Same as in the Fig.~5 except that
the first generation squark masses are presented.}
\end{figure}
\begin{figure}
\caption{Same as in the Fig.~5 except that
the Higgs masses are presented.}
\end{figure}
\begin{figure}
\caption{Same as in the Fig.~5 except that
the chargino and neutralino masses are presented.
The horizontal solid line represents the lightest neutralino.}
\end{figure}


\begin{references}

\bibitem{rev} E. Cremmer, S. Ferrara, L. Girardello, and A.
van Proeyen, Nucl. Phys. {\bf B212} (1983) 413;\\
J. Bagger, Nucl. Phys. {\bf B211} (1983) 302.\\  For
reviews, see,\\
H. P. Nilles, Phys. Rep. {\bf 150} (1984) 1;\\
H. E. Haber and G.  Kane, Phys. Rep. {\bf 117} (1985) 75;\\
A. B. Lahanas and D. V. Nanopoulos, Phys. Rep. {\bf 145}
(1987) 1.

\bibitem{kl} V. Kaplunovsky and J. Louis, Phys. Lett. {\bf B306}
(1993) 269.

\bibitem{barbieri} R. Barbieri, J. Louis and M. Moretti, Phys. Lett.
{\bf B312} (1993) 451;\\
J. Lopez, D. V. Nanopoulos and A.
Zichichi, Phys. Lett. {\bf B319} (1993) 451.

\bibitem{nano} J. Ellis, C. Kounnas and D. V. Nanopoulos,
Nucl. Phys. {\bf B241} (1984) 406 and Nucl. Phys. {\bf B247}
(1984) 373.

\bibitem{ckn} K. Choi, J. E. Kim and H. P. Nilles, Phys. Rev. Lett.
{\bf 73} (1994) 1758.

\bibitem{brignole} A. Brignole, L. E. Ibanez and C. Munoz, Nucl. Phys.
{\bf B422} (1994) 125.

\bibitem{zwirner} S. Ferrara, C. Kounnas and F. Zwirner,
preprint CERN-TH.7192/94 (LPTENS-94/12, UCLA/94/TEP13, hep-th/9405188).

\bibitem{lopez} J. L. Lopez and D. V. Nanopoulos, preprint
CERN-TH.7519/94 (CTP-TAMU-60/94, ACT-18/94, hep-ph/9412332).

\bibitem{gcondensation} S. Ferrara, L. Girardello and H. P. Nilles,
Phys. Lett. {\bf B125} (1983) 457;\\
J. P. Derendinger, L. E.
Ibanez and H. P. Nilles, Phys. Lett. {\bf B155} (1985) 65;\\
M. Dine, R. Rohm, N. Seiberg and E. Witten, Phys. Lett. {\bf B156}
 (1985) 55.\\
For reviews, see for example,\\
H. P. Nilles, Int. J. Mod.
Phys. {\bf A5} (1990) 4199; D. Amaldi et al, Phys. Rep. {\bf 162}
(1988) 169.

\bibitem{cm} B. de Carlos and M. Moretti, Univ. of Oxford preprint
OUTP-94-16P.

\bibitem{bg} P. Binetruy and M. K. Gaillard, Phys. Lett. {\bf B253}
(1991) 119;\\
B. de Carlos, J. A. Casas and C. Munoz, Phys. Lett.
{\bf B299} (1993) 234.

\bibitem{masiero} G. F. Giudice and A. Masiero, Phys. Lett. {\bf B206}
(1988) 480.

\bibitem{kaplunovsky} V. Kaplunovsky, Nucl. Phys. {\bf B307}
(1988) 145 and Nucl. Phys. {\bf B382} (1992) 436.

\bibitem{rge} K. Inoue, A. Kakuto, H. Komatsu, and
S. Takeshita, Prog. Theo. Phys. {\bf 67} (1982) 1889 and
{\bf 68} (1983) 927.

\bibitem{mu} J. E. Kim and H. P. Nilles, Phys. Lett. {\bf B138}
(1984) 150;\\
E. J. Chun, J. E. Kim and H. P. Nilles, Nucl. Phys.
{\bf B370} (1992) 105;\\
J. E. Kim and H. P. Nilles, Seoul National Univ.
preprint SNUTP 94/55 (hep-ph 9406296).

\bibitem{bb} See, for example,\\
V. Barger, M. Berger and P.
Ohmann, Phys. Rev. {\bf D49} (1994) 4908.

\end{references}
\end{document}